\documentclass[10pt, journal] {IEEEtran}
\usepackage{cite}
\usepackage{floatrow}
\DeclareFloatFont{normalsize}{\normalsize}
\usepackage{amsmath}
\usepackage{textcomp}
\usepackage{graphicx}%
\usepackage{amsfonts}%
\usepackage{amssymb}
\usepackage{amsthm}  %
\usepackage{subfig}
\usepackage{tabularx}
\usepackage{multirow}
\usepackage{makecell}
\usepackage{adjustbox}
\usepackage{url}

\usepackage{caption}

\title{\huge Stress Classification 
	from ECG Signals
	Using Vision Transformer
}

\author{Zeeshan Ahmad, \textit{Senior Member, IEEE}, Naimul Khan, \textit{Senior Member, IEEE}}

\begin{document}

	\maketitle

\begin{abstract}

Vision Transformers have shown tremendous success in numerous computer vision applications; however, they have not been exploited for stress assessment using physiological signals such as Electrocardiogram (ECG). In order to get the maximum benefit from the vision transformer for multilevel stress assessment, in this paper, we transform the raw ECG data into 2D spectrograms using short time Fourier transform (STFT). These spectrograms are divided into patches for feeding to the transformer encoder. We also perform experiments with 1D CNN and ResNet-18 (CNN model). We perform leave-one-subject-out cross validation (LOSOCV) experiments on WESAD and Ryerson Multimedia Lab (RML) dataset. One of the biggest challenges of LOSOCV based experiments is to tackle the problem of intersubject variability. In this research, we address the issue of intersubject variability and show our success using 2D spectrograms and the attention mechanism of transformer. Experiments show that vision transformer handles the effect of intersubject variability much better than CNN-based models and beats all previous state-of-the-art methods by a considerable margin. Moreover, our method is end-to-end, does not require handcrafted features, and can learn robust representations. The proposed method achieved 71.01\% and 76.7\% accuracies with RML dataset and WESAD dataset respectively for three class classification and 88.3\% for binary classification on WESAD.

\end{abstract} 


\section{Introduction}

Stress is an active and challenging research area in the field of cognitive science, psychology, biomedical signal processing and affective computing. Stress is recognized as one of the top ten social factors of health
discrepancies by organizations such as the World Health Organisation (W.H.O),
the American Psychological Association (A.P.A)~\cite{mahesh2019requirements}.
A physiological stress is a response of a body
elicited by environmental events or conditions called
stressors. Stressors are external or internal stimuli, events or
conditions. Common stressors include physical stressors, environmental stressors, social stressors, and mental stressors~\cite{giannakakis2019review}.
Long-term stress can have negative effects on both mental and
physical human’s health and can lead to chronic diseases such as eternal depression, cardiac arrest, and poor functioning of organisms such as kidneys and lungs~\cite{giordano2007screening}. Thus, early and precise detection of stress is crucial for human's health. 

ECG signal is readily available and is non-invasive in nature, therefore, it is an attractive choice among the physiological signals for stress analysis and recognition. Studies show that ECG is an instrumental signal for multilevel stress analysis~\cite{giannakakis2019review},~\cite{ahmad2020multi},~\cite{he2019real},~\cite{ahmad2023multilevel}.

One of the major challenges of stress classification using ECG data is the effect of high intersubject variability on the performance of classification models. This effect is more evident when we experiment using LOSOCV setting in which models are trained with data from all-but-one subject, and tested on the held out subject. The common reasons for the high intersubject variablility are the unusual behavior of the participants during data collection and the use of uncallibrated sensors~\cite{ahmad2022survey}. In this paper, we take up the issue of high intersubject variability and its effect on stress classification problem and we successfully reduce the effect by using 2D spectrograms and vision transformer.

Transformer was first introduced in~\cite{vaswani2017attention} and was applied to natural language processing
(NLP) tasks where it shows remarkable performance due to its attention mechanism. Inspired from the substantial success of transformer architectures in
the field of NLP, a vision transformer was proposed in~\cite{dosovitskiy2020image} and it achieved state-of-the-art performance on multiple image recognition benchmarks and is being considered as alternative to convolutional neural networks (CNNs) and their high performance models~\cite{krizhevsky2017imagenet},~\cite{he2016deep} for computer vision applications.

Vision Transformers (ViTs), due to their attention mechanism, are more capable than CNNs to learn weaker biases on backgrounds and textures. They
are equipped with stronger inductive biases towards shapes
and structures, which is more consistent with human cognitive traits and therefore ViTs generalize better than CNNs~\cite{zhang2022delving}. This success is due to the attention mechanism which enables transformers to focus more on relevant features during feature extraction process. We also prove this fact both quantitatively and qualitatively in sections~\ref{ sec : Experimental results} and~\ref{ sec : Discussion} for multilevel stress analysis.

In order to get the maximum benefit of attention mechanism of vision transformer, in this research, we use vision transformer for multilevel stress analysis using ECG signal. First, we transform the raw ECG data into 2D spectrograms using STFT and then we use these spectrgrams as input images for the encoder of vision transformer.

To use spectrogram as an input to vision transformer, we split spectrogram 
into patches and provide the sequence of linear embeddings of these patches as an input to a transformer. Spectrogram patches are treated the same way as tokens (words) were used for NLP application in~\cite{vaswani2017attention}. We used a pre-train vision transformer and performed stress classification from spectrograms in supervised fashion. The advantage of using pre-trained transformer is its robustness to overfitting when fine tuning on specific datasets and its ability to generalize well.


The key contributions of the presented work are:

\begin{enumerate}
	\item To the best of our knowledge this is the first paper where vision transformer is used for stress analysis by transforming raw ECG data into spectrograms using STFT and complete quantitative and qualitative analysis is performed on two datasets for three classes and binary stress classification. Furthermore, the comparison between the performance of vision transformer and ResNet-18 (CNN model) is explained through experiments and visualizations for stress classification.
	\item Using attention mechanism of vision transformer and transforming 1D ECG signal to 2D spectrograms, we achieved higher accuracy of stress recognition beating all the previous state-of-the-art by considerable margin. Transformation of ECG from 1D to 2D enables transformer to concentrate more on the most relevant features that are not present in 1D form. 
	\item Through detailed experiments and visualization across the layers of vision trnasformer, we show that the vision transformer, by using its attention mechanism, addresses the intersubject variability much better than CNN and focus only on the relevant features during the feature extraction process. 
\end{enumerate}

The rest of the paper is organized as follows. Section II describes the related works regarding stress analysis using deep learning models. Section III provides detailed elaboration of proposed method by covering all the technical and mathematical aspects. In section IV, we provide detailed experimental analysis, where the aforementioned contributions are analyzed in detail through large number of experiments and comparison with state-of-the-art models. In section V, we discussed our results both quantitatively and qualitatively. Section VI concludes the paper.

	\begin{figure*}
	\centering
	\includegraphics[width=\linewidth]{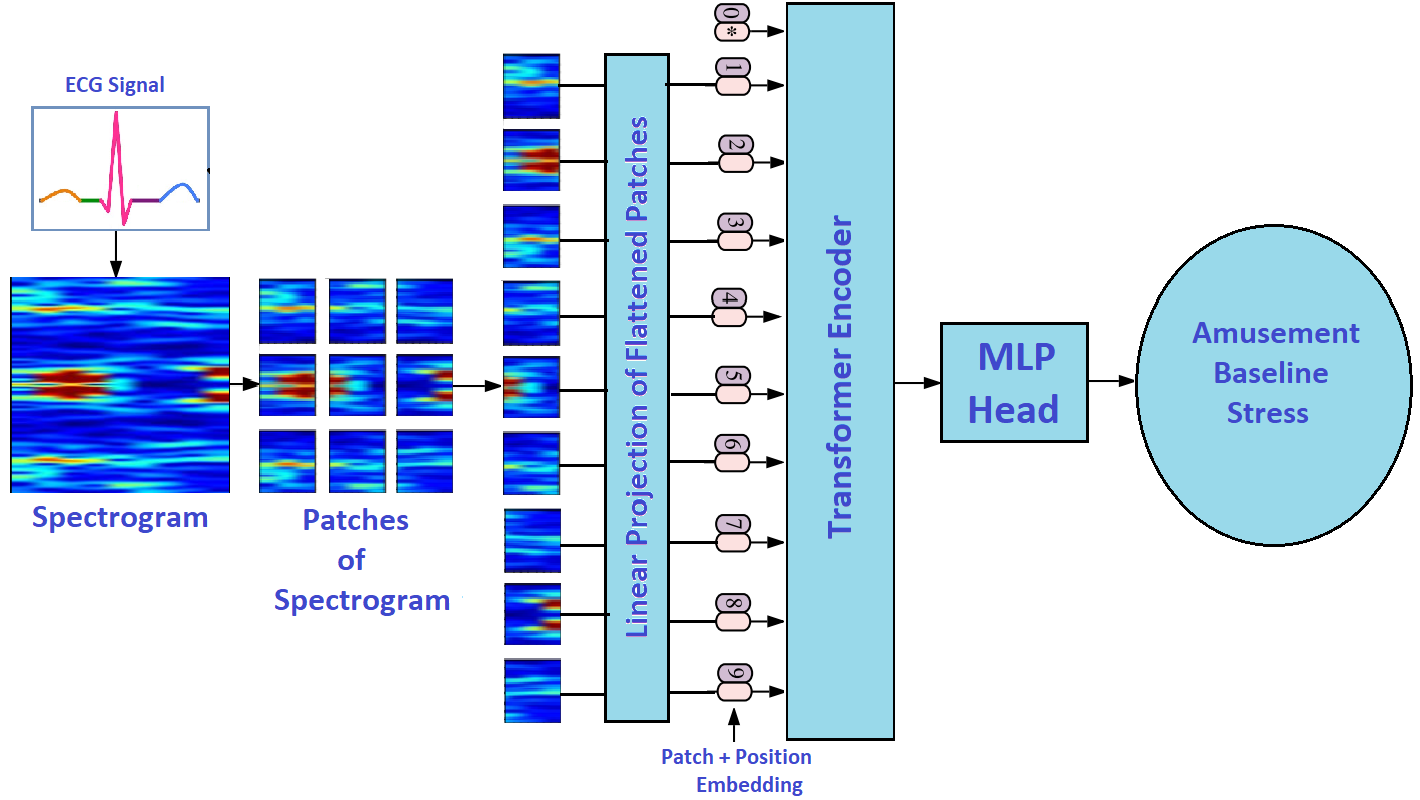}
	\caption{ Overview of the proposed method. We transform an ECG signal into spectrogram which is then split into flattened patches, add position embeddings and then feed to the transformer encoder. Stress recognition is performed by adding an extra learnable “classification token” to the sequence.}
	\label{fig:Overview of proposed method}
\end{figure*}
	
\section{Related Work}

Earlier methods for stress recognition were based on traditional machine learning techniques where the focus was on designing handcrafted features. These traditional machine learning techniques include Hidden Markov Model (HMM), support vector machine (SVM), linear discriminant analysis (LDA), k-means clustering, Gaussian mixture model (GMM), random forest etc~\cite{womack1996classification},~\cite{mcduff2014remote},~\cite{giakoumis2012using},~\cite{kurniawan2013stress},~\cite{tsiknakis2018stress}.

In~\cite{garg2021stress}, various machine learning models are used for the detection of
binary and three class stress using a publicly available multimodal dataset,
WESAD~\cite{schmidt2018introducing}. Sensor data including electrocardiogram (ECG), body temperature (TEMP), respiration (RESP), electromyogram (EMG), and
electrodermal activity (EDA) are used for binary and three class problems. Enough experiments are performed to provide comparison among different machine learning models.
The problems with classical machine learning is that the hand crafted features represents only a subset of the features resulting in difficulty to generalize for unseen data. The other problem was the requirement of domain knowledge about the data.

After the resurgence of neural networks, especially the knock out performance of deep learning models in the field of computer vision and time series classification, the focus of research  shifted towards using deep learning models for stress recognition. In~\cite{huynh2021stressnas}, an optimized deep neural network was proposed for binary and three class stress classification using wrist-worn data from WESAD. Experiments show that proposed method outperforms traditional machine learning methods by considerable margin.
In~\cite{giannakakis2019novel}, a deep learning (DL) multi-kernel
architecture for recognizing stress states through ECG is proposed. The proposed multikernel 1D CNN used 6-fold cross-validation and achieved an state-of-the-art classification accuracy as compared to single kernel CNN.
In~\cite{seo2022deep}, a deep learning architecture is presented  for the accurate
detection of work-related stress using multimodal and heterogeneous signals that were acquired for this study. ECG, RESP, and facial feature data was processed through the DL structure and then feature level and decision level fusions were performed for better stress detection. 
In~\cite{vaitheeshwari2022stress}, machine learning (ML) and deep
learning (DL) models were used for the classification of the physiological data-induced stress conditions in high pressure driving virtual reality (VR) scene. 
A comprehensive analysis was carried out among the significant features of the physiological signal and the raw signals
as well. The observed results reveal that the VR battlefield can effectively arouse stress in humans, and the DL model
can predict the stress condition with good accuracy.
In~\cite{sarkar2020self}, ECG-based emotion recognition problem is solved using self-supervised deep multi-task learning. Four datasets are used in this study to perform emotion recognition. Experiments show that the proposed self supervised approach significantly improves classification performance compared
to a fully-supervised solution.
In~\cite{cho2017deepbreath}, a novel deep learning model called Deep Breath is introduced which
automatically recognises people’s psychological stress level
from their breathing patterns by transforming the patterns into thermal images. These thermal images are used as input to the CNN for binary stress classification on data collected from people exposed to two types of cognitive tasks
(Stroop Colour Word Test, Mental Computation test) with sessions
of different difficulty levels.  
In~\cite{he2019real}, performance comparison of CNN with six
conventional HRV-based methods in acute cognitive stress
detection using only 10s of ECG data, is presented. Stress was induced
by mental calculation for twenty participants. Super-short
temporal windows were used for it was highly desirable
in many practical applications, where the real-time stress
monitoring was an important feature. The results showed
that the performance of CNN was significantly better than
the conventional methods.
A novel end-to-end deep learning architecture based on CNN is proposed in~\cite{cho2019ambulatory} that uses raw ECG signals for stress detection
and validated its performance with two different data sets. Experiments show that the proposed DL models perform better than traditional machine learning models. 

The drawback of the existing deep learning approaches is that they do not take context into account and after training the weights of their layers remain static, no matter the input. Thus to overcome this deficiency, attention-based models are being now used for stress recognition.

In~\cite{duong2021multi}, multimodal stress analysis is performed to classify valence and arousal. Features are extracted from visual, audio, and text modalities using the temporal convolution and
recurrent network with positional embeddings. Finally, late fusion is performed to achieve better performance. In~\cite{yang2022mobile}, a novel end-to-end
emotion recognition system based on a convolution-augmented
transformer architecture is proposed for valence and emotion classification using multiple physiological signals (including blood
volume pulse, electrodermal activity, heart rate, and skin temperature) collected by smart mobile.
Experimental results demonstrate that the presented approach outperforms the
baselines and achieves either state-of-the-art or competitive performance and also generalize well on other datasets. In~\cite{arjun2021introducing}, two vision transformers are used for binary emotion recognition i.e valence and arousal from EEG data. The first method utilizes 2-D images generated
through continuous wavelet transform (CWT) of the raw EEG
signals and the second method directly operates on the raw
signal. The proposed approaches report  high
accuracies for valence and arousal classification on bench mark dataset. In~\cite{behinaein2021transformer}, a deep neural network based on convolutional layers and a transformer mechanism is presented to perform binary classification of stress using
ECG signals. LOSOCV based experiments are performed on 
two publicly available datasets Our experiments show that the proposed model
achieves strong results, comparable or better than the state-of-the-art models for ECG-based stress detection. In~\cite{vazquez2022transformer}, transformer-based self supervised model is proposed to process electrocardiograms (ECG) for emotion recognition. Paper highlighted the fact that the transformers are capable of learning representations
for a signal by giving more importance to relevant parts. Furthermore, the presented approach used self-supervised learning for pre-training from unlabeled data, followed by fine-tuning with labeled data. Experimental results show the dominance of the proposed method. In~\cite{vazquez2021using}, multimodal transformer, using physiological, visual, and audio modalities, is proposed for recognizing and evoking emotions. This research also highlighted the importance of pretraining the transformer using self supervised learning so that it could perform well on downstream task.

\section{Materials and Methods}
	
In this section we explain the proposed method of multilevel stress classification from ECG signal using vision transformer. The overview of the proposed method is shown in Fig.~\ref{fig:Overview of proposed method}. First, we will explain the conversion of ECG signal into spectrograms and then the classification of multilevel stress by spectrograms and vision transformer.

\subsection{ECG signal to spectrogram conversion}\label{sec: ECG to spectrogram}

For using 2D representation as an input to vision transformer, we generate spectrograms from one-second (1s) ECG segments using STFT. The formation of spectrogram from ECG signal is based on STFT below:

\begin{equation}\label{eq : spectrogram}
S_{i}(k,l) = \sum_{m=0}^{M-1}x_{i}(m)w(l-m)e^{-j\frac{2\pi}{M}km}
\end{equation}

where $w(.)$ is the window function, e.g., Hanning window and
$S_{i}(k,l)$ is the STFT of ECG segment ${x_i}$. The spectrogram is then calculated as the square modulus i.e. $|S_{i}(k,l)|^2$.

This equation computes one value STFT, which shows what frequencies are present in a signal at a specific moment in time. To do this, the signal $x_{i}(m)$ is first cut into a small segment using a window function $w(l-m)$. which selects the part of the signal around $l$.
 Then, this windowed segment is multiplied by a complex exponential term, which extracts the frequency information for the frequency bin $k$. By summing all these values from $m=0$ to $M-1$, the equation produces the spectrograms

\begin{figure}[!htbp]
	\centering
	\includegraphics[width=\linewidth]{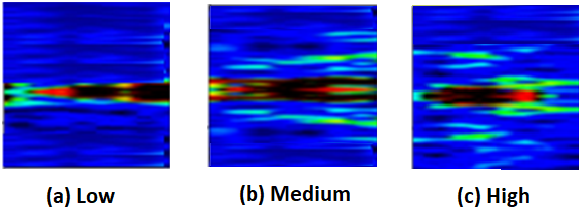}
	\caption{Spectrograms representing low, medium, and high stress levels from the RML dataset. Each spectrogram displays the time–frequency distribution of the ECG signal.}
	\label{fig : spectrogram images}
\end{figure}

As examples, spectrograms of three levels of stress for RML data are shown in Fig.~\ref{fig : spectrogram images}. 

\begin{figure}[!htbp]
	\centering
	\includegraphics[width=0.7\linewidth]{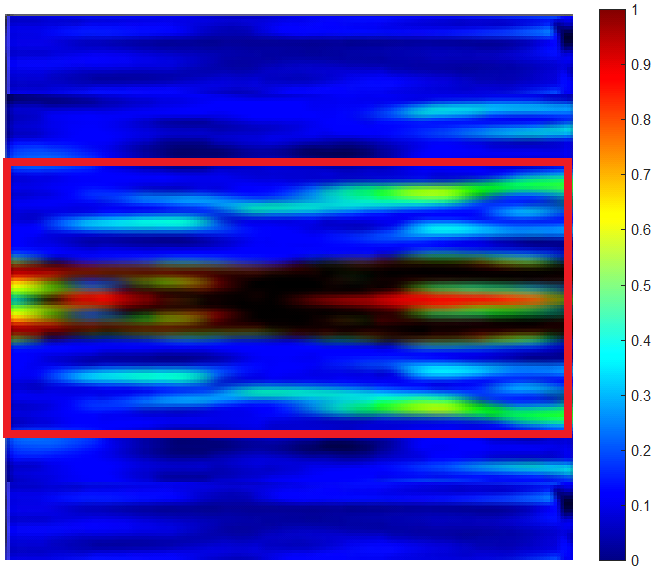}
	\caption{Visualization of features in spectrogram}
	\label{fig : Feature Visualization}
\end{figure}

Fig.~\ref{fig : Feature Visualization} shows the visualization of features in spectrogram. We can see the distribution or the concentration of high intensity features is at the center of the spectrogram and these features fade out as we move away from the center of the images. The location of high intensity features at the center of the spectrogram will help transformer to use its attention mechanism and focus more on the more relevant features for improved performance of stress recognition.

\subsection{Vision Transformer}

The vision transformer used in proposed method consists of transformer encoder and a fully connected classifier. Transformer encoder is pretrained on a large collection of images in a supervised fashion, namely ImageNet-21k, at a resolution of 224 x 224 pixels. It is a BERT (Bidirectional Encoder Representation of Transformer) like encoder introduced for image classification in~\cite{dosovitskiy2020image}.

Images are presented to the model as a sequence of fixed-size patches (16 x 16), which are linearly embedded. [CLS] token is added at the beginning of a sequence to use it for classification tasks. Position embeddings are also added before feeding the sequence to the layers of the transformer encoder.

\begin{figure}[!htbp]
	\centering
	\includegraphics[width= 0.5\linewidth]{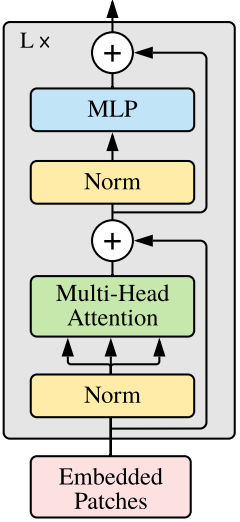}
	\caption{Figure shows one encoder block of a Vision Transformer. This block is repeated $L$ times in the full encoder as indicated by $L_{X}$ at the top.}
	\label{fig : Transformer encoder}
\end{figure}

The transformer encoder contains
alternating layers of Multiheaded Self-Attention (MSA) and
MLP blocks (two layers of MLP with Gaussian Error Linear Unit (GELU) as an activation function). Layer normalization (LN) is applied before every block, and residual connections after every block~\cite{dosovitskiy2020image}. The transformer encoder is shown in Fig.~\ref{fig : Transformer encoder}.

In order to use 2D spectrogram as an input image to vision transformer, we reshape the spectrograms $S \in \mathbf{R}^{H \times W \times C}$ into a
sequence of flattened 2D patches $S_P \in \mathbf{R}^{N \times (P^2 \times C)}$ as shown in Fig.~\ref{fig:Overview of proposed method}.

where,

$(H \times W )$ is the resolution of the original spectrogram, 

$C$ is the number of channels, 

$(P\times P )$ is the resolution of each spectrogram patch, and

$N = HW /P^2$ is the resulting number of patches, which also serves as the effective input sequence length for the transformer. 

For our experiments

$H = W = 224$

$C = 3$

$P = 16$

Finally, before inputting the obtained flattened
patches to the transformer encoder, they are first passed through a trainable linear
projection layer for getting the final patch embeddings and then positional embeddings are added to the patch embeddings for introducing positional information of the
tokens in the sequence as shown in Fig.~\ref{fig:Overview of proposed method}.

\subsection{Feature extraction and classification}

We fine tune the pre-trained vision transformer using the parameters shown in Table.~\ref{tab:parameters for ViT}.
The encoder of vision transformer extracts the most relevant features from input spectrogram images using its attention mechanism and other elements.
Fig.~\ref{fig : Transformer encoder} shows one encoder block of a Vision Transformer. This block is repeated $L$ times in the full encoder as indicated by $L_{X}$ at the top. Each encoder block consists of normalization layer, multi-head self-attention, MLP blocks, and residual connections. The input is normalized at two stages to stabilize training and improve learning~\cite{dosovitskiy2020image}. These useful and informative features extracted by encoder are sent to the classifier for classification.

\begin{table}[h]

	\centering
	\begin{tabular}{c c}
		
		\hline\hline 
		\textbf{Fine Tuned Parameters} & \textbf{Values}   \\\hline\hline
		Learning Rate  &      0.001 \\\hline
		Momentum      &       0.9   \\\hline
		Weight Decay  &      0.005 \\\hline
		BatchSize  &   16 \\
		\hline\hline	
		
	\end{tabular}
		\caption{Parameters for fine tuning the vision transformer}
	\label{tab:parameters for ViT}
\end{table}

The classification task of the proposed methods is multilevel stress classification using ECG signal. 

The classification metrics used for classification are accuracy, precision, recall, and $F_1$ score as shown in Table~\ref{tab : 1D RML Dataset} to Table~\ref{tab: WESAD Binary}. The accuracies, precisions and recalls are calculated using following equations.

\begin{equation}
Accuracy = \frac{TP + TN}{TP + TN + FP + FN}
\end{equation}

\begin{equation}
Precision = \frac{TP}{TP + FP}
\end{equation}

\begin{equation}
Recall = \frac{TP}{TP + FN}
\end{equation}

\begin{equation}
F_1 score = \frac{2 \times(Precision \times  Recall)}{ Precision + Recall}
\end{equation}

where,

$TP$ = True positive 

$TN$ = True negative

$FP$ = False positive

$FN$ = False negative

\section{Experimental Results} \label{ sec : Experimental results}

\subsection{ECG Databases}

Experiments are performed with the Ryerson Multimedia Lab (RML) dataset~\cite{ahmad2023multilevel} and the WESAD dataset~\cite{schmidt2018introducing}. We perform three class classification with RML data while with WESAD dataset, we perform binary as well as three class classification. For both datasets, we perform experiments using leave-one-subject-out cross validation (LOSOCV) setting. In LOSOCV, models are trained with data from all-but-one subject, and tested on the held out subject~\cite{ahmad2022survey}. LOSOCV is more practical than any other cross-validation method especially when the model has to be deployed for real time classification. A model trained and validated on LOSOCV always generalize well on a new subject / patient. 

\subsubsection{RML Dataset}

For RML data collection, 9 subjects were participated and were exposed to varying
stress stimuli while their physiological signals was recorded.
The participants’ electrocardiogram (ECG), galvanic skin
response (GSR) and respiration signals were measured. An ECG sensor, placed around the chest of the subject, was used to record the electrical activity
of the heart. Data was collected at the sampling frequency of 256 Hz. In this research, we only attempted to assess stress levels
utilizing ECG. Utilizing just ECG creates a more practical
solution, since it opens up the possibilities of utilizing even
commercial wearable devices such as the Apple watch. There are three categories of stress in RML dataset. These are Low Stress, Medium Stress and High Stress.

\paragraph{Baseline Experiments}

We performed baseline experiments with RML dataset to validate the effect of
our proposed method shown in Fig.~\ref{fig:Overview of proposed method}. First baseline experiment is performed using 1D CNN on raw ECG data. We used one-second (1s)
window to generate 1D snippets from raw ECG data. We
trained ID CNN on these snippets. The input to ID CNN
is a snippet of size 1 x 256. Five kernels of size 1 x 5 are used
in first convolutional layer, followed by 1 x 2 subsampling
layer. The second convolutional layer contains 10 filters of
same size followed by 1 x 2 subsampling layer. The third
convolutional layer contains 10 filters of size 1 x 4 followed
by 1 x 2 subsampling layer, a fully connected layer and a
classification layer.
The results with 1D CNN are shown in Table~\ref{tab : 1D RML Dataset}.

\begin{table}[h]
	\begin{adjustbox}{width=\columnwidth,center}
		\renewcommand\arraystretch{1}
		\scalebox{0.8}{
			\begin{tabular}{c c c c c}	
				\hline\hline
				\textbf{Testing Sub} & \textbf{\makecell{Accuracy}} & \textbf{\makecell{Precision}} & \textbf{\makecell{Recall}} & \textbf{$F_1$ Score} \\\hline\hline
				2 & 58.9 & 60.9 & 58.9 & 58.6 \\\hline
				3 & 65.4 & 66.4 & 65.4 & 65.4  \\\hline
				4 & 59.3 & 59.4 & 59.3 & 58.9  \\\hline
				5 & 64.7 & 64.2 & 64.7 & 63.2  \\\hline
				11 & 45.3 & 48.8 & 45.3 & 38  \\\hline
				12 & 69.6 & 72.6 & 69.6 & 69.1  \\\hline
				13& 70.6 & 76.3 & 70.6 & 70.3  \\\hline
				14& 66.7 & 69 & 66.7 & 67.2  \\\hline
				16 & 64.8 & 67.2 & 64.8 & 64.1 \\\hline
				\textbf{Average} & \textbf{62.8} & \textbf{62.5} & \textbf{62.8} & \textbf{61.6} \\\hline\hline
				
		\end{tabular}}
	\end{adjustbox}
	\caption{Performance metrics with 1DCNN on RML dataset}
	\label{tab : 1D RML Dataset}
\end{table} 

We also performed experiments with time series transformer on RML Dataset. 

The experimental results are shown in Table~\ref{tab : Time Series Transformer on RML Dataset}

\begin{table}[h]
	\begin{adjustbox}{width=\columnwidth,center}
		\renewcommand\arraystretch{1}
		\scalebox{0.8}{
			\begin{tabular}{c c c c c}	
				\hline\hline
				\textbf{Testing Sub} & \textbf{\makecell{Accuracy}} & \textbf{\makecell{Precision}} & \textbf{\makecell{Recall}} & \textbf{$F_1$ Score} \\\hline\hline
				2 & 61.5 & 61.7 & 61 & 61.8 \\\hline
				3 & 67.18 & 67.83 & 66.94 & 66.95  \\\hline
				4 & 57.67 & 57.47 & 57.56 & 56.84  \\\hline
				5 & 59.94 & 59.45 & 59.65 & 58.84  \\\hline
				11 & 40.6 & 36.9 & 38.68 & 37.47  \\\hline
				12 & 58.68 & 59.78 & 58.31 & 58.51  \\\hline
				13& 60.93 & 60.98 & 60.82 & 60.85  \\\hline
				14& 53.4 & 54.3 & 53.12 & 53.22  \\\hline
				16 & 57.18 & 56.18 & 56.57 & 55.8 \\\hline
				\textbf{Average} & \textbf{57.45} & \textbf{57.18} & \textbf{56.96} & \textbf{56.7} \\\hline\hline
				
		\end{tabular}}
	\end{adjustbox}
	\caption{Performance metrics with time series transformer on RML dataset}
	\label{tab : Time Series Transformer on RML Dataset}
\end{table} 

The poor performance of 1D CNN and time series transformer on raw ECG data compelled us to change the strategy. Thus, we transform raw ECG data to 2D spectrograms as described in section~\ref{sec: ECG to spectrogram}. We used ResNet-18~\cite{he2016deep} to train on spectrograms and experimental results are shown in Table~\ref{tab : Spectrograms RML ResNet}.

\begin{table}[h]
	\begin{adjustbox}{width=\columnwidth,center}
		\renewcommand\arraystretch{1}
		\scalebox{0.8}{
			\begin{tabular}{c c c c c}	
				\hline\hline
				\textbf{Testing Sub} & \textbf{\makecell{Accuracy}} & \textbf{\makecell{Precision}} & \textbf{\makecell{Recall}} & \textbf{$F_1$ Score} \\\hline\hline
				2 & 72.1 & 74.87 & 72.1 & 72.25 \\\hline
				3 & 71.3 & 75.6 & 71.25 & 71.17  \\\hline
				4 & 57.2 & 69.3 & 57.2 & 56.06  \\\hline
				5 & 62.5 & 63.3 & 62.5 & 62.54  \\\hline
				11 & 49.3 & 37.6 & 49.3 & 39.35  \\\hline
				12 & 61.11 & 63.48 & 61.11 & 61.25  \\\hline
				13& 76.1 & 83.7 & 76.15 & 76.22  \\\hline
				14& 64 & 66.13 & 64 & 63.59  \\\hline
				16 & 70.3 & 73.28 & 70.3 & 70.8 \\\hline
				\textbf{Average} & \textbf{65.34} & \textbf{66.2} & \textbf{65.34} & \textbf{63.7} \\\hline\hline
				
		\end{tabular}}
	\end{adjustbox}
	\caption{Performance metrics with ResNet-18 on spectrograms of RML dataset}
	\label{tab : Spectrograms RML ResNet}
\end{table} 

\paragraph{Experiments with the proposed method}

The proposed method is shown in Fig.~\ref{fig:Overview of proposed method}. We fine tuned the pretrained vision transformer with the parameters shown in Table~\ref{tab:parameters for ViT}. The experimental results are shown in Table~\ref{tab: Spectrograms RML ViT}.

\begin{table}[h]
	\begin{adjustbox}{width=\columnwidth,center}
		\renewcommand\arraystretch{1}
		\scalebox{0.8}{
			\begin{tabular}{c c c c c}	
				\hline\hline
				\textbf{Testing Sub} & \textbf{\makecell{Accuracy}} & \textbf{\makecell{Precision}} & \textbf{\makecell{Recall}} & \textbf{$F_1$ Score} \\\hline\hline
				2 & 73  & 76 & 73  & 73.5   \\\hline
				3 & 76.45  & 82.4 & 76.45   & 76.3    \\\hline
				4 &75.14  & 81.6 & 75.14 &74.4   \\\hline
				5 & 69.5  & 69.7 & 69.5  & 69.5  \\\hline
				11 &51  & 62.4 & 51 & 44.7  \\\hline
				12 & 72.9 & 78.7 & 72.9 & 73.3   \\\hline
				13& 75.5 & 82.5  & 75.5  & 75.5  \\\hline
				14& 71.4 & 74.5  & 71.4  & 70.7  \\\hline
				16 & 74.2  & 79.1  & 74.2  & 74.4\\\hline
				\textbf{Average} & \textbf{71.01} & \textbf{76.32} & \textbf{71.01} & \textbf{70.5} \\\hline\hline
				
		\end{tabular}}
	\end{adjustbox}
	\caption{Performance metrics with vision transformer on spectrograms of RML dataset}
	\label{tab: Spectrograms RML ViT}
\end{table}

\subsubsection{WESAD Dataset}

WESAD is a multimodal
dataset recorded with a chest-worn sensor recording ECG,
accelerometer, EMG, respiration, and body temperature and
a wrist-worn sensor recording PPG, accelerometer, electrodermal activity, and body temperature. The data were collected from 17 subjects; each took part in a 2-hour section. Unfortunately, due to device malfunction, data from subject 1 and
12 were discarded. For our experiments, we only
used raw data obtained from the chest-worn ECG signal. In this research, experiments on this dataset were carried
out for detecting and differentiating three affective states
(Amusement, Baseline, Stress) and binary classification (Stress, No-stress).

\paragraph{Baseline Experiments}

We perform baseline experiments on WESAD dataset using ResNet-18. We train the ResNet-18 on the spectrograms of three classes (Amusement, Baseline, Stress). The experimental results are shown in Table~\ref{tab : Spectrograms WESAD ResNet}. 

\begin{table}[h]
	\begin{adjustbox}{width=\columnwidth,center}
		\renewcommand\arraystretch{1}
		\scalebox{0.8}{
			\begin{tabular}{c c c c c}	
				\hline\hline
				\textbf{Testing Sub} & \textbf{\makecell{Accuracy}} & \textbf{\makecell{Precision}} & \textbf{\makecell{Recall}} & \textbf{$F_1$ Score} \\\hline\hline
				2 & 75.1 & 74.75 & 75.1 & 73.88 \\\hline
				3 & 74.34 & 79.75 & 74.34 & 74.72  \\\hline
				4 & 76.46 & 80 & 76.46 & 76.74  \\\hline
				5 & 79.44 & 81.66 & 79.44 & 79.71  \\\hline
				6 & 69.2 & 70.1 & 69.2 & 69.38 \\\hline
				7 & 72.95 & 75.64 & 72.95 & 73.15  \\\hline
				8 & 69.2 & 72.81 & 69.2 & 69.9  \\\hline
				9 & 65.4 & 69.24 & 65.4 & 66.2  \\\hline
				10 & 63.69 & 66.36 & 63.69 & 64.5  \\\hline
				11 & 82.5 & 83.73 & 82.5 & 82.42  \\\hline
				13 & 75.33 & 78.64 & 75.33 & 75.38  \\\hline
				14& 82 & 83.14 & 84 & 82  \\\hline
				15& 66.6 & 67.7 & 66.6 & 66.5  \\\hline
				16 & 82.85 & 85.81 & 82.85 & 82.74 \\\hline
				17 & 85 & 85.34 & 85 & 84.68 \\\hline
				\textbf{Average} & \textbf{74.9} & \textbf{77.54} & \textbf{75.2} & \textbf{75.3} \\\hline\hline
				
		\end{tabular}}
	\end{adjustbox}
	\caption{Performance metrics with ResNet-18 on spectrograms of WESAD data for three class classification}
	\label{tab : Spectrograms WESAD ResNet}
\end{table} 

\paragraph{Experiments with the proposed method}

For validating the performance of our proposed method, we experiments with vision transformer using the spectrograms obtained from WESAD dataset and using the training parameters shown in Table~\ref{tab:parameters for ViT}. The experimental results are shown in Table~\ref{tab : Spectrograms WESAD ViT}. 

These results shown in Tables~\ref{tab: Spectrograms RML ViT} and~\ref{tab : Spectrograms WESAD ViT} highlighted the fact that the visions transformer outperforms ResNet-18 on both datasets. The results are discussed in detail in section~\ref{ sec : Discussion}.

The comparison of the results obtained from WESAD dataset using proposed method with the previous state-of-the-art is shown in Table~\ref{tab:comparison 3class wesad}. We can see that the proposed method beats the previous state-of-the-art with considerable margin.

\paragraph{Binary classification with WESAD dataset}

We also performed binary classification (Stress, No-stress) with WESAD dataset using the proposed method. The experimental results for binary classification and the comparison with the previous state-of-the-art are shown in Tables~\ref{tab: WESAD Binary} and~\ref{tab:comparison binary wesad} respectively. Even for the binary classification, we see in Table~\ref{tab:comparison binary wesad} that the proposed method beats the previous state-of-the-art convincingly.

\begin{table}[h]
	\begin{adjustbox}{width=\columnwidth,center}
		\renewcommand\arraystretch{1}
		\scalebox{0.8}{
			\begin{tabular}{c c c c c}	
				\hline\hline
				\textbf{Testing Sub} & \textbf{\makecell{Accuracy}} & \textbf{\makecell{Precision}} & \textbf{\makecell{Recall}} & \textbf{$F_1$ Score} \\\hline\hline
				2 & 76.22 & 78.5  & 76.2  & 76.1 \\\hline
				3 & 75.5 & 78.8 & 75.5 & 75.8  \\\hline
				4 & 77.14  & 79.11 & 77.14  & 77.4  \\\hline
				5 & 83.83 & 84.5 & 83.83 & 84.03  \\\hline
				6 & 72.7 & 73.4 & 72.7 &72.1  \\\hline
				7 & 73.1 & 75.9 & 73.1 & 73.85  \\\hline
				8 & 73.3 & 76.1 & 73.3 & 73.1  \\\hline
				9 & 69.8 & 72.7 & 69.8 & 70.4  \\\hline
				10 & 66.6  & 68.8 & 66.6 & 66.6  \\\hline
				11 & 83.5 & 84.7 & 83.5 & 83.3  \\\hline
				13 & 75.7 & 79.01 & 75.7 & 75.5  \\\hline
				14& 82.7 & 84 & 82.7 & 82.6  \\\hline
				15 & 70.4  & 71.9  & 70.4  &  70.3 \\\hline
				16 & 84  & 87.3  &84  & 83.9  \\\hline
				17 &85.9  &86.7  & 85.9 & 85.7 \\\hline
				\textbf{Average} & \textbf{76.7} & \textbf{78.8} & \textbf{76.7} & \textbf{76.8} \\\hline\hline
				
		\end{tabular}}
	\end{adjustbox}
	\caption{Performance metrics with Vision Transformer on spectrograms of WESAD data for three class classification}
	\label{tab : Spectrograms WESAD ViT}
\end{table}

\begin{table}[h]
	\begin{adjustbox}{width=\columnwidth,center}
		\renewcommand\arraystretch{1}
		\scalebox{0.8}{
			\begin{tabular}{c c c c c}	
				\hline\hline
				\textbf{Testing Sub} & \textbf{\makecell{Accuracy}} & \textbf{\makecell{Precision}} & \textbf{\makecell{Recall}} & \textbf{$F_1$ Score} \\\hline\hline
				2 & 87.43 & 87.53  & 87.43 & 87.42  \\\hline
				3 & 89.3  & 89.7 & 89.3 &89.3   \\\hline
				4 & 86.5   & 86.6 & 86.5  & 86.4 \\\hline
				5 & 89.1 & 89.7 & 89.1 & 89.1  \\\hline
				6 & 85.9 & 85.9 & 85.9 & 85.9 \\\hline
				7 & 90.9 & 91.2 & 90.9 & 90.9  \\\hline
				8 & 87 & 87.8 & 87  &86.9   \\\hline
				9 & 84.9  & 84.9  & 84.9  & 84.9   \\\hline
				10 & 83.7  & 84.1  & 83.7 & 83.6  \\\hline
				11 & 91.5  & 91.8  & 91.5  & 91.4  \\\hline
				13 & 87.4  & 87.6  & 87.4 & 87.4  \\\hline
				14 & 91.4  & 91.8  & 91.4 & 91.3  \\\hline
				15 & 84.8  & 85.1  & 84.8  & 84.8   \\\hline
				16 & 91.1  & 91.3  & 91.1 & 91.1  \\\hline
				17 & 93.4  & 93.6  & 93.4  & 93.4  \\\hline
				\textbf{Average} & \textbf{88.3} & \textbf{88.6} & \textbf{88.3} & \textbf{88.4} \\\hline\hline
				
		\end{tabular}}
	\end{adjustbox}
	\caption{Performance metrics with Vision Transformer on spectrograms of WESAD data for binary classification}
	\label{tab: WESAD Binary}
\end{table}

\begin{table}[!htbp]
	\centering
	\setlength{\tabcolsep}{3pt}
	\begin{tabular}{c c c}
		
		\hline\hline
		\textbf{Methods} & \textbf{Accuracy} & \textbf{$F_1$ Score} \\\hline\hline 
		P.Schmidt et al.~\cite{schmidt2018introducing}   &  66.29 & 56.03  \\\hline
		P.Garg et al.~\cite{garg2021stress}   &  67.56 & 65.73  \\\hline
		Z.Ahmad et al.~\cite{ahmad2023multilevel}  &  72.7 & 73.1  \\\hline
		
		\textbf{Proposed Method}  & \textbf{76.7} & \textbf{76.8} \\
		\hline\hline			 		
	\end{tabular}
	
	\caption{Performance comparison of the proposed method On the WESAD dataset with other methods for three class classification using LOSOCV splitting.}
	\label{tab:comparison 3class wesad}
\end{table}

\begin{table}[!htbp]
	\centering
	\setlength{\tabcolsep}{3pt}
	\begin{tabular}{c c c}
		
		\hline\hline
		\textbf{Methods} & \textbf{Accuracy} & \textbf{$F_1$ Score} \\\hline\hline
		B.Behinaein et al.~\cite{behinaein2021transformer}   &  80.4 & 69.7  \\\hline
		P.Garg et al.~\cite{garg2021stress}   &  84.17 & 83.34  \\\hline 
		P.Schmidt et al.~\cite{schmidt2018introducing}   &  85.44 & 81.31  \\\hline
	    Bota et al.~\cite{bota2020emotion}  &  85.7 & -  \\\hline
				
		\textbf{Proposed Method}  & \textbf{88.3} & \textbf{88.4} \\
		\hline\hline			 		
	\end{tabular}
	
	\caption{Performance comparison of the proposed method on the WESAD dataset with other methods for binary classification using LOSOCV splitting.}
	\label{tab:comparison binary wesad}
\end{table}

\begin{figure*}[!h]
	\centering
	\includegraphics[width=0.6\linewidth]{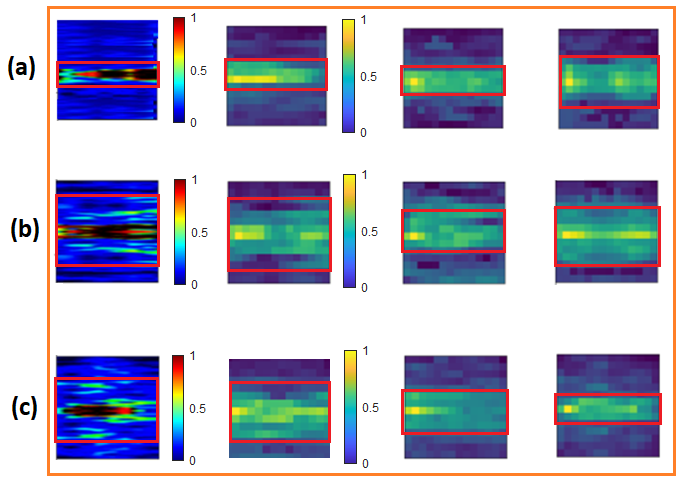}
	\caption{Visualization of attention maps for three stress levels. (a) Low stress, (b) Medium stress, and (c) High stress, each showing the spectrogram with attention maps extracted from the 1st, 5th, and 10th encoder layers.}
	\label{fig : att map Visualization}
\end{figure*}

\begin{figure*}[!h]
	\centering
	\includegraphics[width=0.6\linewidth]{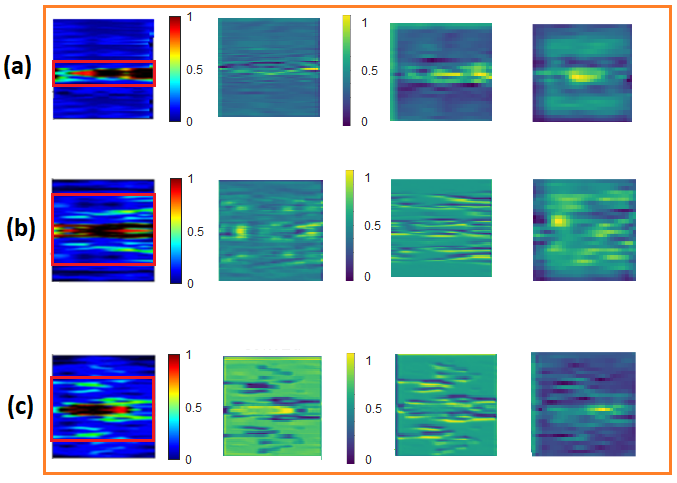}
	\caption{Visualization of Class Activation Maps (CAMs) for three stress levels: (a) low, (b) medium, and (c) high. Each spectrogram is shown with CAMs extracted from the 2nd, 8th, and 16th convolutional layers of ResNet-18.}
	\label{fig : CAMs Visualization}
\end{figure*}

\begin{figure*}
	\vspace{0.5cm}
	\centering
	\includegraphics[width=0.8\linewidth]{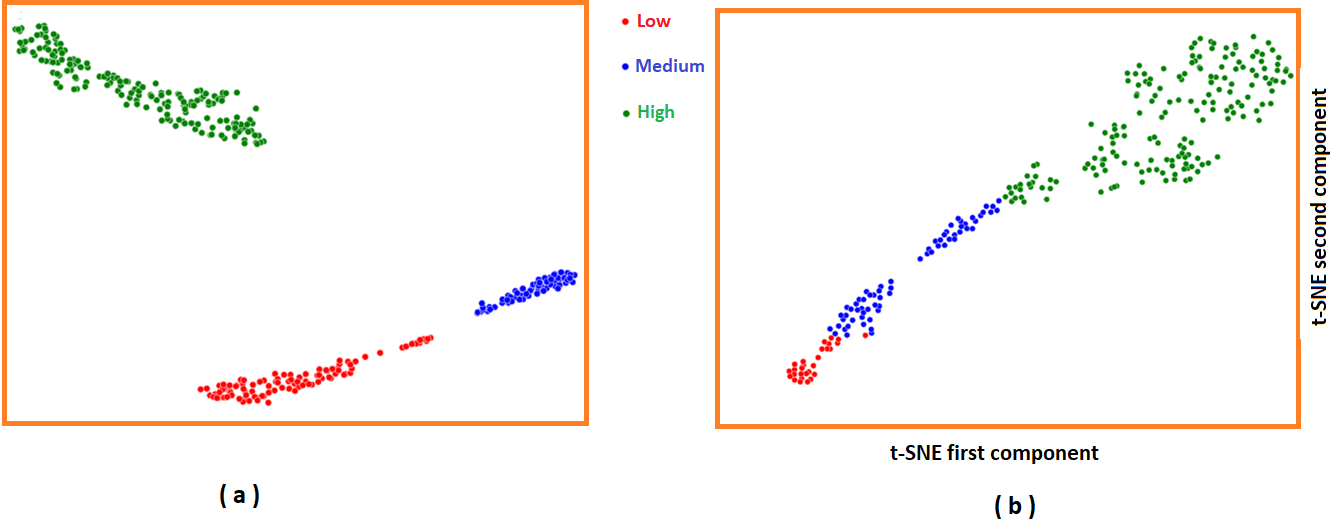}
	\caption{t-SNE visualization of extracted features for Subject 4 of the RML dataset.
		(a) Vision Transformer
		(b) ResNet-18}

	\label{fig : tSNE Visualization}
\end{figure*}

\section{Discussion}\label{ sec : Discussion}

\subsection{Quantitative Analysis}

Tables~\ref{tab:comparison 3class wesad},~\ref{tab:comparison binary wesad} show that the proposed method beats previous state-of-the-art methods for both three class and binary classification. Furthermore, Tables~\ref{tab : 1D RML Dataset},~\ref{tab : Spectrograms RML ResNet},~\ref{tab: Spectrograms RML ViT} and Tables~\ref{tab : Spectrograms WESAD ResNet},~\ref{tab : Spectrograms WESAD ViT} clearly indicate that the vision transformer outperforms ResNet-18 (CNN Model) and 1D CNN by convincing margin on both RML and WESAD datasets. 

The reason of this outstanding performance of vision transformer (ViT) as compared to CNN is that during training, ViT learn stronger inductive biases towards shapes
and structures, which is more consistent with human cognitive traits, while learn weaker biases on backgrounds and textures and thus generalize better than CNNs. Furthermore,  ViTs strengthen these biases and thus gradually
narrow the independent
identically distributed (IID)  and out of distribution (OOD) generalization gaps, especially in
the case of corruption shifts and background shifts. Thus, we can say that ViTs are better at diminishing the effect of local changes.~\cite{zhang2022delving},~\cite{chen2022vision}. Tables~\ref{tab : 1D RML Dataset},~\ref{tab : Spectrograms RML ResNet},~\ref{tab: Spectrograms RML ViT} and tables~\ref{tab : Spectrograms WESAD ResNet},~\ref{tab : Spectrograms WESAD ViT} clearly show that the ResNet-18 fails to address the problem of inter-subject variability of the data. However, transformer addressed this issue well and hence perform well on every subject.

\subsection{Qualitative Analysis}

In Figures~\ref{fig : att map Visualization},~\ref{fig : CAMs Visualization} and~\ref{fig : tSNE Visualization}, we provide feature visualizations using attention maps, class activation maps and t-SNE~\cite{van2008visualizing} to compare the performance of vision transformer and ResNet-18. We extract the attention maps for Low, Medium and High Stress categories from the first, fifth and tenth layer of encoder as shown in Fig.~\ref{fig : att map Visualization}(a), (b), and (c). To provide comparison with ResNet-18, we extract the class activation maps (CAMs) for Low, Medium and High stress categories from the second, eighth and sixteenth convolutional layer of ResNet-18 as shown in Fig.~\ref{fig : CAMs Visualization}(a), (b), and (c). From attention maps, it is clear that the transformer encoder, using its attention mechanism, concentrate more on the features with higher intensity and more relevance. From class activation maps shown in Fig.~\ref{fig : CAMs Visualization}(a), (b), and (c), we can see that, due to lack of attention mechanism, ResNet-18 could not focus entirely on relevant features and also tried to capture irrelevant background features. Furthermore, it is also observed that the CAMs extracted from ResNet-18 for High stress class are bit better than Low and Medium stress. This also shows that ResNet-18 requires more feature distributed input image to perform well which is not the case for vision transformer. Thus, the vision transformer addresses the intersubject variability issue much better than CNN. This can be seen for all Low, Medium and High Stress spectrogram images. 

Fig.~\ref{fig : tSNE Visualization}(a) and (b) shows the visualization of feature grouping of subject 4 of RML dataset using vision transformer and ResNet-18 respectively. It can be clearly observed that the feature grouping with vision transformer is of more qualitative value than with ResNet-18. The features are tightly coupled with their categories in Fig.~\ref{fig : tSNE Visualization}(a) than (b). In Fig.~\ref{fig : tSNE Visualization}(b), the clusters of Medium and High stress are formed at more than one places which shows that the features in the similar categories are loosely bound with each other. Thus, the vision transformer shows better classification performance than ResNet-18.

\section{Conclusion}

In this paper, we used vision transformer for stress assessment using ECG. We transform ECG signal into 2D spectrograms for feeding to the vision transformer. We
also addresses the issue of intersubject variability and through experiments and vision representation, we prove that the vision transformer, using its attention mechanism, handles the effect of intersubject variability much better than CNNs model and
thus beats all previous state-of-the-art methods with considerable
margin. We perform leave-one-subject-out cross validation experiments on WESAD and Ryerson Multimedia Lab (RML) dataset to prove the validity of the proposed method. In our future work, we are planning to do stress classification using raw ECG with time series transformer and design a multimodal fusion framework that could fuse features from vision transformer and time series transformer.

	 
\bibliographystyle{IEEEtran}

\begin{thebibliography}{10}
	\providecommand{\url}[1]{#1}
	\csname url@samestyle\endcsname
	\providecommand{\newblock}{\relax}
	\providecommand{\bibinfo}[2]{#2}
	\providecommand{\BIBentrySTDinterwordspacing}{\spaceskip=0pt\relax}
	\providecommand{\BIBentryALTinterwordstretchfactor}{4}
	\providecommand{\BIBentryALTinterwordspacing}{\spaceskip=\fontdimen2\font plus
		\BIBentryALTinterwordstretchfactor\fontdimen3\font minus
		\fontdimen4\font\relax}
	\providecommand{\BIBforeignlanguage}[2]{{%
			\expandafter\ifx\csname l@#1\endcsname\relax
			\typeout{** WARNING: IEEEtran.bst: No hyphenation pattern has been}%
			\typeout{** loaded for the language `#1'. Using the pattern for}%
			\typeout{** the default language instead.}%
			\else
			\language=\csname l@#1\endcsname
			\fi
			#2}}
	\providecommand{\BIBdecl}{\relax}
	\BIBdecl
	
	\bibitem{mahesh2019requirements}
	B.~Mahesh, T.~Hassan, E.~Prassler, and J.-U. Garbas, ``Requirements for a
	reference dataset for multimodal human stress detection,'' in \emph{2019 IEEE
		International Conference on Pervasive Computing and Communications Workshops
		(PerCom Workshops)}, 2019, pp. 492--498.
	
	\bibitem{giannakakis2019review}
	G.~Giannakakis, D.~Grigoriadis, K.~Giannakaki, O.~Simantiraki, A.~Roniotis, and
	M.~Tsiknakis, ``Review on psychological stress detection using biosignals,''
	\emph{IEEE Transactions on Affective Computing}, vol.~13, no.~1, pp.
	440--460, 2019.
	
	\bibitem{giordano2007screening}
	M.~Giordano, P.~Tirelli, T.~Ciarambino, A.~Gambardella, N.~Ferrara,
	G.~Signoriello, G.~Paolisso, and M.~Varricchio, ``Screening of depressive
	symptoms in young--old hemodialysis patients: Relationship between beck
	depression inventory and 15-item geriatric depression scale,'' \emph{Nephron
		Clinical Practice}, vol. 106, no.~4, pp. c187--c192, 2007.
	
	\bibitem{ahmad2020multi}
	Z.~Ahmad and N.~M. Khan, ``Multi-level stress assessment using multi-domain
	fusion of ecg signal,'' in \emph{2020 42nd Annual International Conference of
		the IEEE Engineering in Medicine \& Biology Society (EMBC)}, 2020, pp.
	4518--4521.
	
	\bibitem{he2019real}
	J.~He, K.~Li, X.~Liao, P.~Zhang, and N.~Jiang, ``Real-time detection of acute
	cognitive stress using a convolutional neural network from
	electrocardiographic signal,'' \emph{IEEE Access}, vol.~7, pp.
	42\,710--42\,717, 2019.
	
	\bibitem{ahmad2023multilevel}
	Z.~Ahmad, S.~Rabbani, M.~R. Zafar, S.~Ishaque, S.~Krishnan, and N.~Khan,
	``Multilevel stress assessment from ecg in a virtual reality environment
	using multimodal fusion,'' \emph{IEEE Sensors Journal}, vol.~23, no.~23, pp.
	29\,559--29\,570, 2023.
	
	\bibitem{ahmad2022survey}
	Z.~Ahmad and N.~Khan, ``A survey on physiological signal-based emotion
	recognition,'' \emph{Bioengineering}, vol.~9, no.~11, p. 688, 2022.
	
	\bibitem{vaswani2017attention}
	A.~Vaswani, N.~Shazeer, N.~Parmar, J.~Uszkoreit, L.~Jones, A.~N. Gomez,
	L.~Kaiser, and I.~Polosukhin, ``Attention is all you need,'' \emph{Advances
		in Neural Information Processing Systems}, vol.~30, 2017.
	
	\bibitem{dosovitskiy2020image}
	A.~Dosovitskiy, L.~Beyer, A.~Kolesnikov, D.~Weissenborn, X.~Zhai,
	T.~Unterthiner, M.~Dehghani, M.~Minderer, G.~Heigold, S.~Gelly \emph{et~al.},
	``An image is worth 16x16 words: Transformers for image recognition at
	scale,'' \emph{International Conference on Learning Representations (ICLR)},
	2021.
	
	\bibitem{krizhevsky2017imagenet}
	A.~Krizhevsky, I.~Sutskever, and G.~E. Hinton, ``Imagenet classification with
	deep convolutional neural networks,'' \emph{Communications of the ACM},
	vol.~60, no.~6, pp. 84--90, 2017.
	
	\bibitem{he2016deep}
	K.~He, X.~Zhang, S.~Ren, and J.~Sun, ``Deep residual learning for image
	recognition,'' in \emph{Proceedings of the IEEE Conference on Computer Vision
		and Pattern Recognition}, 2016, pp. 770--778.
	
	\bibitem{zhang2022delving}
	C.~Zhang, M.~Zhang, S.~Zhang, D.~Jin, Q.~Zhou, Z.~Cai, H.~Zhao, X.~Liu, and
	Z.~Liu, ``Delving deep into the generalization of vision transformers under
	distribution shifts,'' in \emph{Proceedings of the IEEE/CVF Conference on
		Computer Vision and Pattern Recognition}, 2022, pp. 7277--7286.
	
	\bibitem{womack1996classification}
	B.~D. Womack and J.~H. Hansen, ``Classification of speech under stress using
	target driven features,'' \emph{Speech Communication}, vol.~20, no. 1--2, pp.
	131--150, 1996.
	
	\bibitem{mcduff2014remote}
	D.~McDuff, S.~Gontarek, and R.~Picard, ``Remote measurement of cognitive stress
	via heart rate variability,'' in \emph{2014 36th Annual International
		Conference of the IEEE Engineering in Medicine and Biology Society}, 2014,
	pp. 2957--2960.
	
	\bibitem{giakoumis2012using}
	D.~Giakoumis, A.~Drosou, P.~Cipresso, D.~Tzovaras, G.~Hassapis, A.~Gaggioli,
	and G.~Riva, ``Using activity-related behavioural features towards more
	effective automatic stress detection,'' \emph{PLoS ONE}, 2012.
	
	\bibitem{kurniawan2013stress}
	H.~Kurniawan, A.~V. Maslov, and M.~Pechenizkiy, ``Stress detection from speech
	and galvanic skin response signals,'' in \emph{Proceedings of the 26th IEEE
		International Symposium on Computer-Based Medical Systems}, 2013, pp.
	209--214.
	
	\bibitem{tsiknakis2018stress}
	M.~Tsiknakis, ``Stress detection from speech using spectral slope
	measurements,'' \emph{Pervasive Computing Paradigms for Mental Health}, vol.
	207, p.~41, 2018.
	
	\bibitem{garg2021stress}
	P.~Garg, J.~Santhosh, A.~Dengel, and S.~Ishimaru, ``Stress detection by machine
	learning and wearable sensors,'' in \emph{26th International Conference on
		Intelligent User Interfaces-Companion}, 2021, pp. 43--45.
	
	\bibitem{schmidt2018introducing}
	P.~Schmidt, A.~Reiss, R.~Duerichen, C.~Marberger, and K.~Van~Laerhoven,
	``Introducing wesad, a multimodal dataset for wearable stress and affect
	detection,'' in \emph{Proceedings of the 20th ACM International Conference on
		Multimodal Interaction}, 2018, pp. 400--408.
	
	\bibitem{huynh2021stressnas}
	L.~Huynh, T.~Nguyen, T.~Nguyen, S.~Pirttikangas, and P.~Siirtola, ``Stressnas:
	Affect state and stress detection using neural architecture search,'' in
	\emph{Adjunct Proceedings of the 2021 ACM International Joint Conference on
		Pervasive and Ubiquitous Computing}, 2021, pp. 121--125.
	
	\bibitem{giannakakis2019novel}
	G.~Giannakakis, E.~Trivizakis, M.~Tsiknakis, and K.~Marias, ``A novel
	multi-kernel 1d convolutional neural network for stress recognition from
	ecg,'' in \emph{2019 8th International Conference on Affective Computing and
		Intelligent Interaction Workshops and Demos (ACIIW)}, 2019, pp. 1--4.
	
	\bibitem{seo2022deep}
	W.~Seo, N.~Kim, C.~Park, and S.-M. Park, ``Deep learning approach for detecting
	work-related stress using multimodal signals,'' \emph{IEEE Sensors Journal},
	2022.
	
	\bibitem{vaitheeshwari2022stress}
	R.~Vaitheeshwari, S.-C. Yeh, E.~H.-K. Wu, J.-Y. Chen, and C.-R. Chung, ``Stress
	recognition based on multiphysiological data in high-pressure driving vr
	scene,'' \emph{IEEE Sensors Journal}, vol.~22, no.~20, pp. 19\,897--19\,907,
	2022.
	
	\bibitem{sarkar2020self}
	P.~Sarkar and A.~Etemad, ``Self-supervised ecg representation learning for
	emotion recognition,'' \emph{IEEE Transactions on Affective Computing}, 2020.
	
	\bibitem{cho2017deepbreath}
	Y.~Cho, N.~Bianchi-Berthouze, and S.~J. Julier, ``Deepbreath: Deep learning of
	breathing patterns for automatic stress recognition using low-cost thermal
	imaging in unconstrained settings,'' in \emph{2017 Seventh International
		Conference on Affective Computing and Intelligent Interaction (ACII)}, 2017,
	pp. 456--463.
	
	\bibitem{cho2019ambulatory}
	H.-M. Cho, H.~Park, S.-Y. Dong, and I.~Youn, ``Ambulatory and laboratory stress
	detection based on raw electrocardiogram signals using a convolutional neural
	network,'' \emph{Sensors}, vol.~19, no.~20, p. 4408, 2019.
	
	\bibitem{duong2021multi}
	A.-Q. Duong, N.-H. Ho, H.-J. Yang, G.-S. Lee, and S.-H. Kim, ``Multi-modal
	stress recognition using temporal convolution and recurrent network with
	positional embedding,'' in \emph{Proceedings of the 2nd Multimodal Sentiment
		Analysis Challenge}, 2021, pp. 37--42.
	
	\bibitem{yang2022mobile}
	K.~Yang, B.~Tag, Y.~Gu, C.~Wang, T.~Dingler, G.~Wadley, and J.~Goncalves,
	``Mobile emotion recognition via multiple physiological signals using
	convolution-augmented transformer,'' 2022.
	
	\bibitem{arjun2021introducing}
	A.~Arjun, A.~S. Rajpoot, and M.~R. Panicker, ``Introducing attention mechanism
	for eeg signals: Emotion recognition with vision transformers,'' in
	\emph{2021 43rd Annual International Conference of the IEEE Engineering in
		Medicine \& Biology Society (EMBC)}, 2021, pp. 5723--5726.
	
	\bibitem{behinaein2021transformer}
	B.~Behinaein, A.~Bhatti, D.~Rodenburg, P.~Hungler, and A.~Etemad, ``A
	transformer architecture for stress detection from ecg,'' in \emph{2021
		International Symposium on Wearable Computers}, 2021, pp. 132--134.
	
	\bibitem{vazquez2022transformer}
	J.~Vazquez-Rodriguez, G.~Lefebvre, J.~Cumin, and J.~L. Crowley,
	``Transformer-based self-supervised learning for emotion recognition,'' in
	\emph{2022 26th international conference on pattern recognition (ICPR)},
	2022, pp. 2605--2612.
	
	\bibitem{vazquez2021using}
	J.~Vazquez-Rodriguez, ``Using multimodal transformers in affective computing,''
	in \emph{2021 9th International Conference on Affective Computing and
		Intelligent Interaction Workshops and Demos (ACIIW)}, 2021, pp. 1--5.
	
	\bibitem{bota2020emotion}
	P.~Bota, C.~Wang, A.~Fred, and H.~Silva, ``Emotion assessment using feature
	fusion and decision fusion classification based on physiological data: Are we
	there yet?'' \emph{Sensors}, vol.~20, no.~17, p. 4723, 2020.
	
	\bibitem{chen2022vision}
	X.~Chen, C.-J. Hsieh, and B.~Gong, ``When vision transformers outperform
	resnets without pre-training or strong data augmentations,'' \emph{ICLR
		Conference Paper}, 2022.
	
	\bibitem{van2008visualizing}
	L.~Van~der Maaten and G.~Hinton, ``Visualizing data using t-sne,''
	\emph{Journal of Machine Learning Research}, vol.~9, no.~11, 2008.
	
\end{thebibliography}

\begin{IEEEbiography}[{\includegraphics[width=0.8in,height=1in,clip]{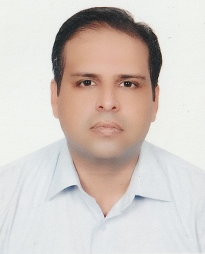}}]{Dr. Zeeshan Ahmad}  (senior member, IEEE) is as an Assistant Professor at University of Niagara Falls Canada. He earned his B.Eng in Electrical Engineering from NED University of Engineering and Technology, Karachi, Pakistan in 2001 and M.Sc in Electrical Engineering from National University of Sciences and Technology (NUST), Pakistan in 2005.  Zeeshan joined Department of Electrical, Computer and Biomedical Engineering at Toronto Metropolitan University in 2015 and earned his M.Eng in 2017 and Ph.D. in 2021 with specialization in Deep Learning for Computer Vision. From 2021 to 2024, Zeeshan worked as Postdoctoral Fellow at Toronto Metropolitan University. His teaching and research include Data Analytics, Deep Learning for Computer Vision and NLP, Multimodal Machine Learning, and Physiological Signal Processing. 
\end{IEEEbiography}

\begin{IEEEbiography}[{\includegraphics[width=0.8in,height=1in,clip]{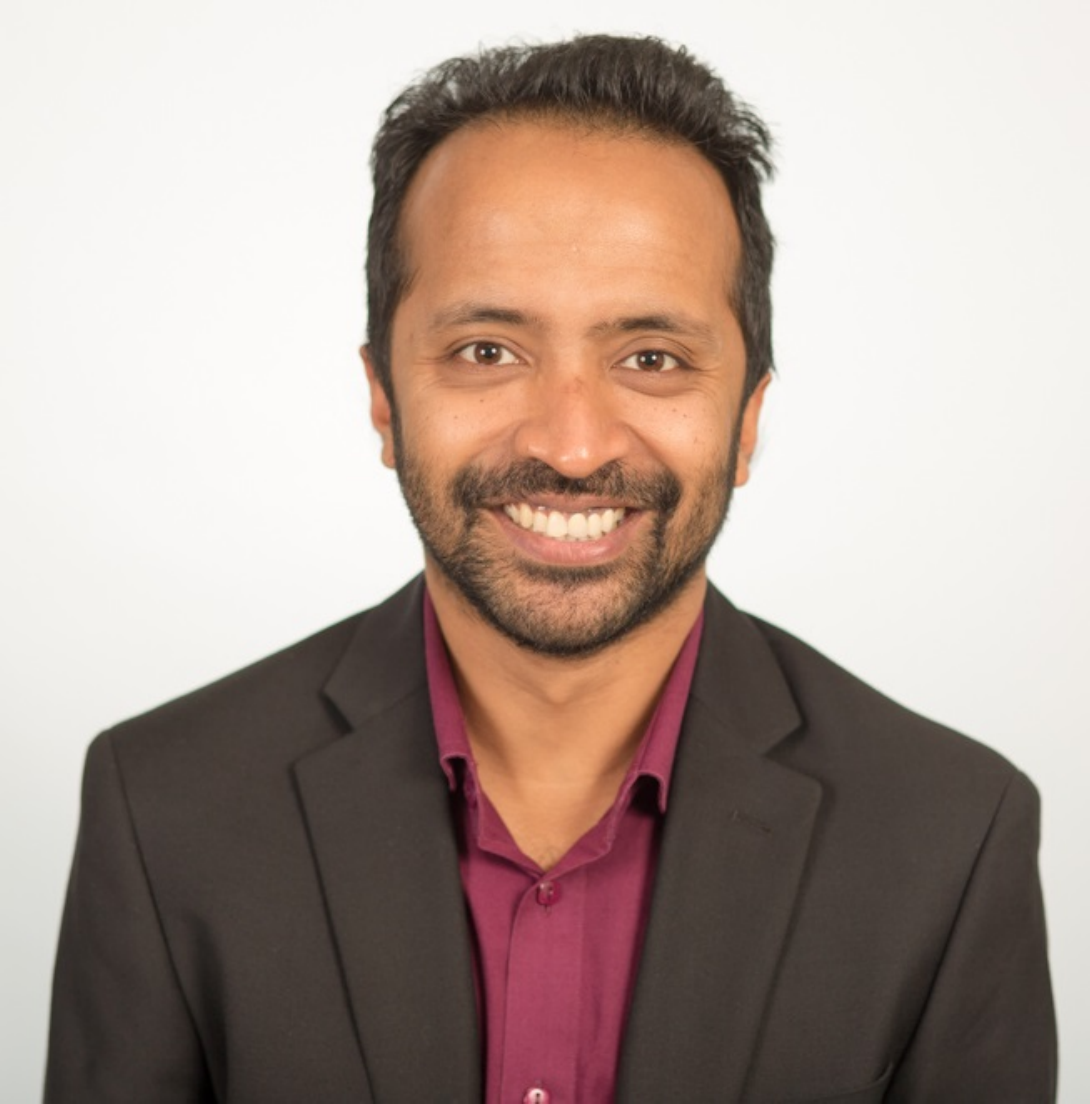}}]{Dr. Naimul Khan} (senior member, IEEE) is an associate professor in Electrical and Computer Engineering at Toronto Metropolitan University with a cross-appointment at the creative school (Digital Media). He is the director of the TMU Multimedia Research Laboratory. His interdisciplinary work spans artificial intelligence, physiological signal processing, and immersive technologies, with applications in health, accessibility, and human-computer interaction. He leads projects that integrate multimodal data (such as EEG, ECG, and video) with machine learning to develop real-time, personalized systems for emotion recognition, digital mental health, and therapeutic VR. 
	
\end{IEEEbiography}

\end{document}